\documentclass[11pt]{amsart}
\usepackage[margin=1.1in]{geometry}
\usepackage{amsmath,amssymb,amsthm,mathtools,enumitem,xcolor}
\usepackage{lmodern}
\usepackage[T1]{fontenc}
\usepackage{microtype}
\usepackage[colorlinks=true,linkcolor=blue!50!black,citecolor=blue!50!black,urlcolor=blue!50!black]{hyperref}

\newtheorem{theorem}{Theorem}[section]
\newtheorem{proposition}[theorem]{Proposition}
\newtheorem{lemma}[theorem]{Lemma}

\newtheorem{conjecture}[theorem]{Conjecture}
\theoremstyle{definition}

\theoremstyle{remark}
\newtheorem{remark}[theorem]{Remark}

\newcommand{\N}{\mathbb N}
\newcommand{\Z}{\mathbb Z}
\newcommand{\PP}{\mathsf P}
\newcommand{\FO}{\mathsf{FO}}
\DeclareMathOperator{\Seq}{Seq}
\DeclareMathOperator{\Rep}{Rep}
\DeclareMathOperator{\pr}{pr}

\title{On Presburger Arithmetic Expanded with Prime Numbers}
\author{Joris Nieuwveld}
\address{Department of Computer Science, University of Oxford}
\date{November 2026}

\begin{document}

\begin{abstract}
We prove unconditionally that the first-order theory of Presburger arithmetic expanded by a predicate for the primes is undecidable. Bateman, Jockusch and Woods proved this in 1993 assuming Dickson's conjecture. We use the counter-machine framework of Karimov, Nieuwveld and Ouaknine and their idea of using `arbitrary permutations', and the number-theoretic input comes from Hildebrand and Maier's 1988 proof that the limit points of $k$ consecutive normalised prime gaps have positive measure. We use a discrete version, which they obtain before limits are taken. The key lemma is that although Hildebrand and Maier's result does not seem to give arbitrary permutations of consecutive prime gaps, it gives arbitrary permutations for the differences $p_i - q_i$ for two finite blocks of consecutive primes $p_0 < \cdots < p_k$ and $q_0 < \cdots < q_k$.
Moreover, with the same key lemma, we show that Presburger arithmetic expanded with the function that maps $n$ to the $n$th prime defines multiplication.
\end{abstract}

\maketitle

\section{Introduction}
Presburger arithmetic, the first-order theory of $\langle\N;+\rangle$, denoted $\FO(\N; +)$, is well known to be decidable. 
Naturally, the question arose of which seemingly weaker versions of multiplication can be added to Presburger arithmetic to obtain undecidability and which define multiplication. 
For instance, expanding Presburger arithmetic with a predicate for divisibility, coprimality, or the set of squares makes multiplication definable; see the survey of B\`es~\cite{Bes} for these and other examples.
One obvious candidate remained open: Is access to knowing whether a number is prime sufficient to establish undecidability? Can one define multiplication?

Bateman, Jockusch, and Woods \cite{BJW} studied two expansions of Presburger arithmetic by the predicate $\PP$ of primes using Dickson's conjecture, which is also sometimes referred to as the linear case of Schinzel's Hypothesis H.
\begin{conjecture}[Dickson's conjecture]
    Let $a_1,\dots,a_k \ge 1$ and $b_1,\dots,b_k \in \Z$ such that for every prime number $p$, there is a number $x$ such that none of $a_1x+b_1,\dots,a_kx+b_k$ is divisible by $p$. 
    Then there are infinitely many $n$ such that $a_1n+b_1,\dots,a_kn+b_k$ are simultaneously prime. 
\end{conjecture}
The conjecture is strong: it implies the twin prime conjecture. 
Bateman, Jockusch, and Woods show that $\FO(\N;<,\PP)$ is decidable assuming Dickson's conjecture (and even more strongly, that the monadic second-order theory of this expansion would be decidable). 
An unconditional proof of this decidability would effectively settle the twin prime conjecture.
Moreover, they showed that Dickson's conjecture implies that multiplication is definable in $\langle\N; +, \PP\rangle$, which implies the theory is undecidable.
They showed that for any $N$, there are numbers $a$ and $c$ such that $ax+c+0,ax+c+2,ax+c+6,\dots,ax+c+N^2+N$ satisfies the premise of the conjecture while no number in between is ever prime. 
Hence, one gets access to the image of a quadratic polynomial and thus multiplication.

Assuming Dickson's conjecture, Kaplan and Shelah~\cite{KS} showed that $\FO(\Z;0,1,+,\PP\cup\{-p:p\in\PP\})$ is decidable and does not define order. As order is implicit in Presburger arithmetic via $x\le y\iff\exists z\in\N\colon x+z=y$, order seems essential for undecidability. For instance, Goldbach's conjecture is easily expressed in $\FO(\N;+,\PP)$, whereas, in Kaplan and Shelah's structure, one can naturally express only a weaker version: every even number is a sum or difference of two primes.

An unconditional proof of the undecidability of $\FO(\N; +, \PP)$ has remained open; the problem is discussed by B\`es~\cite{Bes}, C\'egielski, Richard, and Vsemirnov~\cite{CRV}, Kaplan and Shelah~\cite{KS}, and Karimov, Nieuwveld, and Ouaknine~\cite{KNO}.
We show undecidability in this note. 
\begin{theorem}\label{thm:main set}
The theory $\FO(\N;+,\PP)$ is undecidable.
\end{theorem}
Despite establishing undecidability, we do not obtain a definition of multiplication. 
We also study the theory $\FO(\N;+,\pr)$, where $\pr(n)$ is the $n$th prime, whose decidability was questioned in~\cite{Bes,CRV}.
So $\pr(1) = 2$. To make $\pr$ total, we set $\pr(0) = 0$, but this choice does not matter.
\begin{theorem}\label{thm:main function}
Multiplication is definable in the structure $\langle \N;+,\pr\rangle$.
\end{theorem}

\subsection*{Intuition of the proof.}
We prove Theorem~\ref{thm:main set} using the recent framework of Karimov, Nieuwveld and Ouaknine \cite{KNO}: it is enough to show that a structure \emph{simulates counter machines}. 
A counter machine is a (Minsky) machine with finitely many states and counters that can be incremented, decremented and tested for zero, whose halting problem is undecidable in general.
A structure simulates counter machines if every counter machine can be translated into a first-order sentence that holds if and only if the machine halts.
Halting asserts the existence of an encoded \emph{trace} of the halting computation, i.e., the successive configurations of the machine for which initial and halting configurations and a transition relation must be first-order expressible.
Karimov et al. informally call structures that simulate counter machines \emph{rich} as they need to realise every suitable pattern of some sort.
Although the primes are expected to behave quite randomly within the set of natural numbers, controlling exact prime configurations remains difficult. 

A first attempt would be to look at the results on arithmetic progressions in primes of Green and Tao~\cite{GT10}.
Kaplan and Shelah~\cite[Propostion 3.6]{KaplanShelah2017} note that it implies that for any set $S \subset \{1,\dots,k\}$, there are infinitely many numbers $a, b$ such that for $0 \le i \le k$, $ai+b$ is prime if and only if $i \in S$. 
That is, the primes contain every finite pattern.
However, we need to multiply by the parameter $a$, which is impossible in Presburger arithmetic. 

A second attempt would be to look at consecutive prime gaps and presume that they have \emph{arbitrary permutations}. 
That is, for any $k$ and permutation $\sigma$ of $\{1,\dots,k\}$, there is a finite sequence of consecutive primes $p_0 < \cdots < p_k$ such that $p_{\sigma(1)} - p_{\sigma(1)-1} < \dots < p_{\sigma(k)} - p_{\sigma(k)-1}$. 
In this case, we would only need inequalities and subtraction, which are definable in $\langle \N;+,\PP\rangle$, and the argument of~\cite{KNO} would apply almost verbatim.

Unfortunately, it is not known whether consecutive prime gaps have arbitrary permutations. 
There are two kinds of results in this direction.
First, monotone patterns occur infinitely often~\cite{BFT}, but we are not aware of any generalisation to any other fixed pattern. 
Secondly, there is a result by Hildebrand and Maier.
\begin{theorem}[Hildebrand--Maier~\cite{HM}]\label{thm:HM continuous}
    Let $k \ge 1$. Then there is a constant $c_k > 0$ such that for all sufficiently large $T$, the set of limit points of
    \begin{equation*}
        \left\{\left(\frac{\pr(n+1)-\pr(n)}{\log(n)},\dots,\frac{\pr(n+k)-\pr(n+k-1)}{\log(n)} \right) : n \in \N_{\ge 1}\right\} \cap [0, T]^k
    \end{equation*}
    has Lebesgue measure at least $c_kT^k$. 
\end{theorem}
If $c_k$ were larger than $1-1/k!$, consecutive prime gaps would have arbitrary permutations.
Although $c_k$ is not explicitly computed, it appears to be much smaller, and so we cannot conclude arbitrary permutations.
Differences of two limit points would allow arbitrary permutations, but the two limit points may arise at unrelated scales $\log(n)$. 
We therefore work with the discrete count (Theorem~\ref{thm:HM}), where any two \emph{runs}, i.e., finite blocks of consecutive primes, can be compared directly.
\begin{lemma}[Two-run lemma]\label{lem:tworun}
For every $k\ge1$ and every permutation $\sigma$ of $\{1,\dots,k\}$ there are two runs
\[p_0<p_1<\dots<p_k\qquad\text{and}\qquad q_0<q_1<\dots<q_k\]
with the following property. Put $e_i=p_i-q_i$ and $\Delta=\max_{1\le i\le k}|e_i - e_0|$. Then
\begin{enumerate}[label=\textup{(\alph*)}]
\item $e_{\sigma(1)}<e_{\sigma(2)}<\dots<e_{\sigma(k)}$, and
\item every gap of the second run exceeds $2\Delta$: $q_j-q_{j-1}>2\Delta$ for $1\le j\le k$.
\end{enumerate}
\end{lemma}
From this lemma, item (a) gives arbitrary permutations while item (b) allows us to determine whether a given $p_i$ and $q_j$ have the same index and thus whether their difference is used in the encoding.
We make this precise in the proof of Theorem~\ref{thm:main set} in Section~\ref{sec:logic}.

The functional version of the result, Theorem~\ref{thm:main function}, follows similarly.
Consider two runs $(\pr(n+i))_{i=1}^k$ and $(\pr(r+i))_{i=1}^k$, and let $\psi(i) := \pr(n+1)-\pr(r+1) \ge \pr(n+i) - \pr(r+i)$. 
By Lemma~\ref{lem:tworun}, there are $r$ and $n$ such that for $1 \le i \le k$, $\psi(i)$ holds exactly when $i$ is a square. 
Conversely, if $\psi(4)$ holds and $i_3 = 2i_2 - i_1 + 2$ for all $0 < i_1< i_2 < i_3$ that are consecutive numbers satisfying $\psi$, any number satisfying $\psi$ is a square.
Hence, multiplication is definable.

\section{From two runs to undecidability}\label{sec:logic}

We prove Theorem~\ref{thm:main set} using the counter-machine framework of~\cite{KNO} while assuming Lemma~\ref{lem:tworun}.

\subsection{Counter-machine simulation}\label{sec:counter machines}
We recall the notion from \cite[\S1.2]{KNO}, specialised to index tuples of length one. A structure $\mathcal{M}$ with domain $\N$ \emph{simulates counter machines} if there are $\ell\ge1$, maps $\Seq\colon\N^\ell\to\N^*$ and $\Rep\colon\N^\ell\to\N^*$ (the encoded sequence and the elements representing its terms, respectively), and formulas $\mathrm{rep}(\mathbf{x},y)$, $\mathrm{cnst}_K(\mathbf{x},y)$ (computable from $K$), $\mathrm{succ}(\mathbf{x},y,z)$, $\mathrm{inc}(\mathbf{x},y,z)$ and $\mathrm{eq}(\mathbf{x},y,z)$, such that for all $\mathbf{x}\in\N^\ell$:
\begin{enumerate}[label=(\arabic*),leftmargin=2.2em]
\item $\Seq$ is surjective onto the finite sequences of natural numbers;
\item $|\Seq(\mathbf{x})|=|\Rep(\mathbf{x})|$, and the entries of $\Rep(\mathbf{x})$ are distinct;
\item Representation: $\mathrm{rep}(\mathbf{x},y)$ holds if and only if $y$ occurs in $\Rep(\mathbf{x})$;
\item Constant value $K$: $\mathrm{cnst}_K(\mathbf{x},y)$ holds if and only if $y=\Rep(\mathbf{x})_i$ for some $i$ with $\Seq(\mathbf{x})_i=K$;
\item Successor: $\mathrm{succ}(\mathbf{x},y,z)$ holds if and only if $y=\Rep(\mathbf{x})_i$ and $z=\Rep(\mathbf{x})_{i+1}$ for some $i$;
\item Increment: $\mathrm{inc}(\mathbf{x},y,z)$ holds if and only if $y=\Rep(\mathbf{x})_i$, $z=\Rep(\mathbf{x})_j$ and $\Seq(\mathbf{x})_j=\Seq(\mathbf{x})_i+1$;
\item Equality: $\mathrm{eq}(\mathbf{x},y,z)$ holds if and only if $y=\Rep(\mathbf{x})_i$, $z=\Rep(\mathbf{x})_j$ and $\Seq(\mathbf{x})_i=\Seq(\mathbf{x})_j$.
\end{enumerate}

\begin{theorem}[{\cite[Theorem 2]{KNO}}]\label{thm:KNO}
Every structure that simulates counter machines has an undecidable first-order theory.
\end{theorem}

The proof of this theorem translates a given counter machine into a first-order sentence $\varphi$ that simulates its trace. To check that this is indeed a valid trace, one only needs to compare nearby terms in the trace.
If the counter machine halts, this trace is finite, which some $\mathbf{x} \in \N^\ell$ witnesses. 
That is, $\exists\mathbf{x}\varphi(\mathbf{x})$ holds, and for this $\mathbf{x}$, $\Seq(\mathbf{x})$ is exactly the trace of the counter machine and $\Rep(\mathbf{x})$ encodes the corresponding sequence over which one iterates in the first-order formula.

In Presburger arithmetic, we write $x\le y$ for $\exists z\,(x+z=y)$, $x<y$ for $x\le y\wedge x\ne y$, and $x-y$ for the unique $z$ satisfying $x=z+y$ when $y \le x$. Statements such as ``$p$ is a prime in $[a,c]$'' and ``$p,p'$ are consecutive primes'' are thus first-order.

\subsection{Applying counter machines to prime gaps}

We use $\ell=6$ and write $\mathbf{x}=(a,b,c,a',c',\delta)$.
For primes $p\in[a,c]$ and $q\in[a',c']$ write $p\sim_\mathbf{x} q$ if $|(p-q)-(a-a')|\le\delta$, that is,
\[
p+a'\le q+a+\delta \quad\text{and}\quad q+a\le p+a'+\delta .
\]
Let $\mathrm{Valid}(\mathbf{x})$ be the formula stating that $a\le b\le c$ and that for every prime $p\in[a,c]$ there is exactly one prime $q\in[a',c']$ with $p\sim_\mathbf{x}q$.
For valid $\mathbf{x}$ and a prime $p\in[a,c]$ we call this $q$ the \emph{match} of $p$ and write $\mathrm{Match}(p)$.
Both $\mathrm{Valid}$ and the graph of $\mathrm{Match}$ are first-order definable in $\langle\N;+,\PP\rangle$.
Put $v(p)=p-\mathrm{Match}(p)\in\Z$. We use $v$ only in comparisons: $v(p)\le v(p')$ stands for the formula $p+\mathrm{Match}(p')\le p'+\mathrm{Match}(p)$, and similarly for $<$.

\begin{proposition}\label{prop:sim}
If Lemma~\ref{lem:tworun} holds, then $\langle\N;+,\PP\rangle$ simulates counter machines.
\end{proposition}
\begin{proof}
If $\mathrm{Valid}(\mathbf{x})$ fails, let $\Seq(\mathbf{x})$ and $\Rep(\mathbf{x})$ be empty sequences.
Otherwise, call the primes in $(a,b]$ \emph{marks}, let $\Rep(\mathbf{x})$ list the primes in $(b,c]$ in increasing order, and set
\[
\Seq(\mathbf{x})_i=\#\{p'\text{ a mark}: v(p')\le v(\Rep(\mathbf{x})_i)\}.
\]
Writing $p,p'$ for elements of $\Rep(\mathbf{x})$, the formulas are:
\begin{itemize}[leftmargin=2em]
\item $\mathrm{rep}(\mathbf{x},p)$: $\mathrm{Valid}(\mathbf{x})$, $b<p\le c$, and $\PP(p)$;
\item $\mathrm{cnst}_K(\mathbf{x},p)$: $\mathrm{rep}(\mathbf{x},p)$, and exactly $K$ marks $p'$ satisfy $v(p')\le v(p)$;
\item $\mathrm{succ}(\mathbf{x},p,p')$: $\mathrm{rep}(\mathbf{x},p)$, $\mathrm{rep}(\mathbf{x},p')$, $p<p'$, and no prime lies strictly between $p$ and $p'$;
\item $\mathrm{inc}(\mathbf{x},p,p')$: $\mathrm{rep}(\mathbf{x},p)$, $\mathrm{rep}(\mathbf{x},p')$, and exactly one mark $p''$ satisfies $v(p)<v(p'')\le v(p')$;
\item $\mathrm{eq}(\mathbf{x},p,p')$: $\mathrm{rep}(\mathbf{x},p)$, $\mathrm{rep}(\mathbf{x},p')$, and no mark $p''$ satisfies $v(p)<v(p'')\le v(p')$ or $v(p')<v(p'')\le v(p)$.
\end{itemize}
``Exactly $K$'' is a first-order statement computable from $K$.

\emph{Items (2)--(7).} If $\mathrm{Valid}(\mathbf{x})$ fails, every formula is false and both sequences are empty. Otherwise (2)--(5) are immediate. For (6) and (7), let $p=\Rep(\mathbf{x})_i$ and $p'=\Rep(\mathbf{x})_j$. If $v(p)\le v(p')$, then $\Seq(\mathbf{x})_j-\Seq(\mathbf{x})_i$ is the number of marks $p''$ with $v(p)<v(p'')\le v(p')$. If instead $v(p)>v(p')$, $\Seq(\mathbf{x})_j \le \Seq(\mathbf{x})_i$ and no mark $p''$ satisfies $v(p)<v(p'')\le v(p')$. Hence $\mathrm{inc}(\mathbf{x},p,p')$ holds if and only if $\Seq(\mathbf{x})_j=\Seq(\mathbf{x})_i+1$. Likewise $\mathrm{eq}(\mathbf{x},p,p')$ holds if and only if $\Seq(\mathbf{x})_i=\Seq(\mathbf{x})_j$, by the same argument applied to both orders of $p$ and $p'$.

\emph{Item (1).} The empty sequence comes from any $\mathbf{x}$ with $\mathrm{Valid}(\mathbf{x})$ false. Let $(t_1,\dots,t_\tau)$ be non-empty, choose $r\ge\max(1,t_1,\dots,t_\tau)$, and set $k=r+\tau$. Order the positions $1,\dots,k$ by the keys
\[\kappa(j)=j\ \ (1\le j\le r),\qquad \kappa(r+i)=t_i+\tfrac{i}{\tau+1}\ \ (1\le i\le \tau).\]
The keys are distinct, and exactly $t_i$ of the first $r$ positions have a key below $\kappa(r+i)$. Let $\sigma$ list the positions in increasing order of key, and let $p_0<\dots<p_k$ and $q_0<\dots<q_k$ be the runs given by Lemma~\ref{lem:tworun} applied to $\sigma$. Put
\[
\mathbf{x}=(p_0,\,p_r,\,p_k,\,q_0,\,q_k,\,\Delta).
\]
For $0\le i\le k$ we have $p_i\sim_\mathbf{x}q_i$, since $|(p_i-p_0)-(q_i-q_0)|=|e_i-e_0|\le\Delta$. For $j\ne i$ we have $|q_j-q_i|>2\Delta$ by (b), so $|(p_i-p_0)-(q_j-q_0)|>2\Delta-\Delta=\Delta$. Hence $\mathbf{x}$ is valid and $\mathrm{Match}(p_i)=q_i$, so $v(p_i)=e_i$. The marks are $p_1,\dots,p_r$ and $\Rep(\mathbf{x}) = (p_{r+1},\dots,p_k)$. By (a), $e_j\le e_{r+i}$ holds if and only if $\kappa(j)\le\kappa(r+i)$, so $\Seq(\mathbf{x})_i=t_i$.
\end{proof}

\begin{proof}[Proof of Theorem~\ref{thm:main set}, given Lemma~\ref{lem:tworun}]
Combine Proposition~\ref{prop:sim} with Theorem~\ref{thm:KNO}.
\end{proof}

\begin{remark}\label{rem:quantifiers needed predicate}
Inspecting the sentence constructed in the proof of \cite[Theorem~2]{KNO} shows that it has the form $\exists\mathbf{x}\,[\mathrm{Valid}(\mathbf{x})\wedge\psi(\mathbf{x})]$ with $\psi$ a $\Pi_2$-formula. Hence already the $\exists^6\forall^*\exists^*$-fragment of $\FO(\N;+,\PP)$ is undecidable, where the $6$ comes from $\mathbf{x}\in\N^6$. In fact $c'$ can be eliminated: taking $c'=c+a'+\delta-a$ and applying Lemma~\ref{lem:tworun} with $k+1$ in place of $k$, with the extra index placed last in $\sigma$, so that $q_{k+1} - q_k > 2\Delta$, the interval $[a',c']$ contains exactly $q_0,\dots,q_k$ in the proof of Proposition~\ref{prop:sim}, since $c+a'-a=q_k+(e_k-e_0)$ and the gaps of the $q$-run exceed $2\Delta$. This gives the $\exists^5\forall^*\exists^*$-fragment. The number of inner quantifiers in our construction is unbounded, since $\mathrm{cnst}_K$ uses $K+1$ quantifiers of both kinds.
\end{remark}

\section{The \texorpdfstring{$n$th}{nth} prime function and multiplication}

We next prove Theorem~\ref{thm:main function} while assuming Lemma~\ref{lem:tworun}. We first recall a classical result from Putnam~\cite{Put57}. Let $\mathsf{Sq}$ be the predicate for the square numbers.
\begin{lemma}\label{lem:Putnam squares}
    The structure $\langle\N;+,\mathsf{Sq}\rangle$ defines multiplication.
\end{lemma}
\begin{proof}
    First note that $x = y^2$ if and only if $x$ is a square, $x + 2y + 1$ is also a square and no square lies in between.
    Then the identity $(x+y)^2 = x^2+y^2+2xy$ shows that $z = xy$ if and only if $(x+y)^2 = x^2+y^2+2z$. 
\end{proof}
\begin{proof}[Proof of Theorem~\ref{thm:main function}, given Lemma~\ref{lem:tworun}]
By Lemma~\ref{lem:Putnam squares} it suffices to define $\mathsf{Sq}$. The positive squares are the sequence $u_0=1$, $u_1=4$, $u_{j+2}=2u_{j+1}-u_j+2$, which we encode as follows.

For $k,n,r\in\N$ write $\mathbf{x}=(k,n,r)$, and let $\psi_\mathbf{x}(i)$ be the formula
\[
1\le i\le k\;\wedge\;\pr(n+i)+\pr(r+1)\le\pr(n+1)+\pr(r+i).
\]
Let $\varphi(\mathbf{x})$ be the formula
\begin{multline*}
    \psi_{\mathbf{x}}(4)\wedge\forall i_1,i_2,i_3 : i_1<i_2<i_3\rightarrow \\\Big[\Big(\bigwedge_{j=1}^{3}\psi_{\mathbf{x}}(i_j)\wedge\forall i : i_1<i<i_3 \rightarrow\big(\neg\psi_{\mathbf{x}}(i)\vee i=i_2\big)\Big)\rightarrow i_3+i_1=2i_2+2\Big].
\end{multline*}

Note that $\psi_\mathbf{x}(1)$ holds whenever $k\ge1$, and $\varphi(\mathbf{x})$ implies $\psi_\mathbf{x}(4)$. If $i_1<i_2<i_3$ are consecutive among the numbers satisfying $\psi_\mathbf{x}$, then $i_3+i_1=2i_2+2$.

If $\varphi(\mathbf{x})$ holds, then $k\ge4$ and every number that satisfies $\psi_\mathbf{x}$ is a square. Indeed, if $\psi_\mathbf{x}(2)$ held, the next number satisfying $\psi_\mathbf{x}$ would be $5$, contradicting $\psi_\mathbf{x}(4)$. Similarly, $\psi_\mathbf{x}(3)$ cannot hold. By induction along the recurrence, $\psi_\mathbf{x}(i)$ holding means $i$ is a square.

Conversely, let $k\ge4$, and let $\sigma$ list first the squares in $\{2,\dots,k\}$, then $1$, then the non-squares in $\{2,\dots,k\}$. Lemma~\ref{lem:tworun} applied to $\sigma$ gives runs of consecutive primes, so let $p_i=\pr(n+i)$ and $q_i=\pr(r+i)$ for suitable $n,r$. By~(a), $p_i+q_1\le p_1+q_i$ holds exactly for the squares $i\le k$, so $\varphi(k,n,r)$ holds. 

Therefore,
\[
\mathsf{Sq}(i)\iff i=0\;\vee\;\exists n,r\,\big(\varphi(4i,n,r)\wedge\psi_{4i,n,r}(i)\big).\qedhere
\]
\end{proof}

\begin{remark}\label{rem:quantifiers needed function}
    As in Remark~\ref{rem:quantifiers needed predicate}, this method establishes that the $\exists^*\forall^*\exists^*$-fragment of $\FO(\N;+,\pr)$ is undecidable, using the undecidability of Hilbert's tenth problem.
    As many numbers needed for the squares and multiplication in the polynomial are in the outer existential quantifier, Hilbert's tenth problem does not directly bound the number of quantifiers in a block.
    By a result of Jones~\cite{Jon82}, there is an explicit polynomial $f \in \Z[\mathbf{y}, \mathbf{x}]$ such that determining whether, for a given $\mathbf{y}$, there exists $\mathbf{x}$ such that $f(\mathbf{y}, \mathbf{x}) = 0$ is undecidable. As $1$ is existentially definable ($z = 1$ is the only number with $z \ne 2z = \pr(z)$), we can write a component of $\mathbf{y}$ as $1+1+\dots+1$ in a sentence. 
    Hence, there are $A, B, C \in \N$ such that the $\exists^A\forall^B\exists^C$-fragment of $\FO(\N;+,\pr)$ is undecidable, which we do not compute here.
\end{remark}

\section{Proof of Lemma~\ref{lem:tworun}}\label{sec:proof tworun}

We begin with the discrete version of Hildebrand and Maier's Theorem~\ref{thm:HM continuous}.

For a run $p_0<\dots<p_k$, define its \emph{gap vector} as $(p_1-p_0,\dots,p_k-p_{k-1})$. For $x>1$, $T>0$, $N\ge1$, and $\mathbf{n}\in\{1,\dots,N\}^k$, let
\[\mathcal{B}_{\mathbf{n}}=\prod_{j=1}^k\left((n_j-1)\frac{T\log x}{N},\ n_j\frac{T\log x}{N}\right].\]
A box is \emph{hit} if it contains the gap vector of a run of $k+1$ consecutive primes.

\begin{theorem}[\cite{HM}]\label{thm:HM}
For every $k\ge1$ there are $c\in(0,1]$ and $T > 0$ such that for all $N\ge1$ there are arbitrarily large $x$ (depending on $N$) for which at least $cN^k$ of the boxes $\mathcal{B}_{\mathbf{n}}$ are hit.
\end{theorem}


This is precisely inequality (17) in Hildebrand and Maier's paper. To recall the argument, they construct a matrix containing $\gg R$ tuples of $k+1$ consecutive primes (their equation (14)), and show that each of the $N^k$ boxes contains at most $O(R/N^k)$ such tuples (their equation (19)). Hence $\gg N^k$ boxes contain such tuples, proving (17). This holds at a single value of $x$, and for arbitrarily large values of $x$.

The next lemma is a pigeonhole-principle argument showing that, for a given vector $\mathbf{w}$, a large enough subset of $\{1,\dots,N\}^k$ contains two points that differ by a small multiple of $\mathbf{w}$.

\begin{lemma}\label{lem:lines}
Let $\mathbf{w}\in\Z^k\setminus\{\mathbf{0}\}$, $L\ge2$, and $\mathcal{H}\subseteq\{1,\dots,N\}^k$ with $\#\mathcal{H}>N^k/L+\|\mathbf{w}\|_1N^{k-1}$. Then there are $\mathbf{b}\in\mathcal{H}$ and $1\le t\le L-1$ with $\mathbf{b}+t \mathbf{w}\in\mathcal{H}$.
\end{lemma}
\begin{proof}
The lines $\{\mathbf{b}+t\mathbf{w} : t \in \Z\}$ partition the box $\{1,\dots,N\}^k$. Each line $\ell$ meets $\{1,\dots,N\}^k$ in consecutive points $\mathbf{b}_\ell,\mathbf{b}_\ell+\mathbf{w},\dots,\mathbf{b}_\ell+(s_\ell-1)\mathbf{w}$ for some $\mathbf{b}_\ell$ and $s_\ell \ge 1$. 
Cut each such progression into pieces of $L$ consecutive points, the first possibly shorter. As two points in the same piece differ by $t\mathbf{w}$ with $1\le t\le L-1$, it suffices to show there are fewer than $\#\mathcal{H}$ pieces.

The number of pieces is at most $N^k/L$ (at most the number of pieces containing no $\mathbf{b}_\ell$) plus the number of lines meeting $\{1,\dots,N\}^k$ (the number of pieces containing a $\mathbf{b}_\ell$). A point $\mathbf{b}_\ell$ satisfies $\mathbf{b}_\ell-\mathbf{w}\notin \{1,\dots,N\}^k$, so $b_{\ell,j}\le w_j$ for some $j$ with $w_j>0$, or $b_{\ell,j}>N+w_j$ for some $j$ with $w_j<0$. 
This occurs at most $\sum_{j=1}^k |w_j|N^{k-1}=\|\mathbf{w}\|_1N^{k-1}$ times, and so there are at most $N^k/L+\|\mathbf{w}\|_1N^{k-1}<\#\mathcal{H}$ pieces.
\end{proof}

We can now prove Lemma~\ref{lem:tworun}.

\begin{proof}[Proof of Lemma~\ref{lem:tworun}]
Fix $k\ge1$ and a permutation $\sigma$ of $\{1,\dots,k\}$, let $s(i)=\sigma^{-1}(i)$ be the desired rank of $1 \le i \le k$, and set $s(0)=0$.
The idea is to find two runs $p_0<\dots<p_k$ and $q_0 < \dots<q_k$ where we set $e_i = p_i - q_i$ for $0 \le i \le k$. Then we ensure that each $e_i-e_0$ is, up to a common scaling factor, approximately $4ks(i)$, with an error of less than $k$ on the same scale.
That is, the gap vectors $(p_i-p_{i-1})_{i=1}^k$ and $(q_i-q_{i-1})_{i=1}^k$ differ by roughly a multiple of $\mathbf{w}$, where $w_j=4k(s(j)-s(j-1))$.
To find such runs, we apply Lemma~\ref{lem:lines} to the boxes hit in Theorem~\ref{thm:HM}.

Let $c \in (0, 1]$ be as in Theorem~\ref{thm:HM}, $L$ an integer with $L>\frac{4}{c}$, and $N$ an integer with
\[
    \frac{c}{4k} N\ge \max(8k^2L, \|\mathbf{w}\|_1).
\]
Take $T$ and $x$ as in Theorem~\ref{thm:HM}, let $\mathcal{H} \subset \{1,\dots,N\}^k$ be the set of indices $\mathbf{n}$ of hit boxes $\mathcal{B}_{\mathbf{n}}$, let $\widetilde{\mathcal{H}}$ those with $n_j\ge\frac{c}{4k} N+1$ for all $j$, and write $u=\frac{T\log x}{N}$ for the side length of the boxes.
Then, $\#\mathcal{H} \ge cN^k$. 
As $1 < \frac{c}{4k}N$, at most $k \cdot (\frac{c}{4k} N+1)N^{k-1} < \frac{c}{2}N^k$
boxes have some $n_j<\frac{c}{4k} N+1$, so
\[
\#\widetilde{\mathcal{H}} > \#\mathcal{H} - \tfrac{c}{2}N^k \ge \tfrac{c}{4}N^k + \tfrac{c}{4}N^k > N^k/L+\|\mathbf{w}\|_1N^{k-1}\,.
\]

Lemma~\ref{lem:lines} gives $\mathbf{b},\mathbf{b}+t\mathbf{w}\in\widetilde{\mathcal{H}}$ with $1\le t\le L-1$. Take a run $q_0<\dots<q_k$ with gap vector $(q_j-q_{j-1})_{j=1}^k$ in $\mathcal{B}_{\mathbf{b}}$, and a run $p_0<\dots<p_k$ with gap vector $(p_j-p_{j-1})_{j=1}^k$ in $\mathcal{B}_{\mathbf{b}+t\mathbf{w}}$. Since the side-length $u$ boxes are $tw_j = t\cdot 4k(s(j)-s(j-1))$ steps apart in coordinate $j$, we have 
\begin{equation*}
    \big|e_j-e_{j-1}-4tuk\big(s(j)-s(j-1)\big)\big| <u\,.
\end{equation*}
Summing over $j\le i$ gives two telescoping sums, which imply
\[
\big|(e_i-e_0)-4kt\,s(i)\,u\big|<iu\le ku .
\]

If $s(i')=s(i)+1$, then $(e_{i'}-e_0)-(e_i-e_0)>4ktu-2ku\ge 2ku>0$, which proves~(a). For~(b), $\Delta=\max_i|e_i-e_0|<4k^2tu+ku<4k^2Lu\le\frac{c}{8k} Nu$, while every gap of the $q$-run is at least $\frac{c}{4k} Nu>2\Delta$.
\end{proof}

\section{Discussion}

The proof of Theorem~\ref{thm:main set} does not show that multiplication is definable in $\langle\N;+,\PP\rangle$, while Theorem~\ref{thm:main function} shows it is definable in $\langle\N;+,\pr\rangle$.
The key difference is that the functional version allows arithmetic on prime indices. 
Hence, we expect that the techniques from this note cannot settle definability of multiplication in $\langle\N;+,\PP\rangle$, and that one needs more precise information about linear relations between primes such as those coming from Dickson's conjecture. 

By Remarks~\ref{rem:quantifiers needed predicate} and~\ref{rem:quantifiers needed function}, the $\exists^*\forall^*\exists^*$-fragments of both $\FO(\N;+,\PP)$ and $\FO(\N;+,\pr)$ are undecidable. 
We expect that these results are not optimal, and that the $\exists^*\forall^*$-fragments of both theories are also undecidable.
For example, assuming Dickson's conjecture, for every finite sequence $(t_i)_{i=1}^\tau$ of natural numbers, there is a number $n$ such that $\pr(n+2i+1)-\pr(n+2i) = 2t_i+2$ for $1 \le i \le \tau$.
Using the techniques from Section~\ref{sec:counter machines}, we can encode a given counter machine in the $\exists^*\forall^*$-fragment of $\FO(\N;+,\pr)$, giving undecidability. 
Meanwhile, we expect that the existential fragment of $\FO(\N;+,\PP)$ is decidable again assuming Dickson's conjecture.

Although order is implicit in Presburger arithmetic, it is also definable in the structure $\langle\Z; +, \PP\rangle$: By a result of Tao~\cite{Tao14}, every odd number at least 3 is the sum of at most 5 primes, so every non-negative even number is the sum of at most 6 primes. Thus, $x \ge 0$ if and only if $2x$ is the sum of at most 6 primes. Hence, we can recover the natural numbers in the structure $\langle \Z; +, \PP \rangle$, making its first-order theory undecidable. 
Definability of order seems necessary, as Kaplan and Shelah~\cite{KS} showed that $\FO(\Z;0,1,+,\PP\cup\{-p:p\in\PP\})$ is decidable assuming Dickson's conjecture.

\section*{Acknowledgements}
The author was supported by the Glasstone Benefaction, University of Oxford [Violette and Samuel Glasstone Research Fellowships in Science 2025].

\textbf{AI declaration.}
Claude Opus 5.5 was used to prepare this note.
In particular, when prompted with Lemma~\ref{lem:tworun} and Hildebrand and Maier's paper~\cite{HM}, it returned a first draft of the proof presented in Section~\ref{sec:proof tworun}.
The remainder of the proof is our own, although AI helped frame and smooth our construction.
Furthermore, AI was used to proofread this note and suggest improvements.
The author takes full responsibility for the correctness of the results.

\bibliographystyle{amsplain}
\bibliography{bib}
\end{document}